\documentclass[useAMS,usenatbib,usegraphicx]{mn2e}

\usepackage{epsf}
\usepackage{cite}
\usepackage{rotating}

\def\beq{\begin{equation}}
\def\eeq{\end{equation}}

\title[Measuring Reionization using the 21-cm PDF]{Measuring 
the History of Cosmic Reionization using the 21-cm PDF from
Simulations} \author[Ichikawa, Barkana, Iliev, Mellema,
Shapiro]{Kazuhide Ichikawa$^{1,2,3}$, Rennan Barkana$^{1,4,5}$
\thanks{E-mail: barkana@wise.tau.ac.il (RB)}, Ilian T.\ Iliev$^{6,7}$,
Garrelt Mellema$^8$, \and Paul R.\ Shapiro$^9$\\ $^1$ Institute for
Cosmic Ray Research, University of Tokyo, Kashiwa 277-8582, Japan \\
$^2$ Department of Physics and Astronomy, University College London,
Gower Street, London, WC1E 6BT, U.K.\ \\ $^3$ Department of Micro
Engineering, Kyoto University, Kyoto 606-8501, Japan \\ $^4$ Division
of Physics, Mathematics and Astronomy, California Institute of
Technology, Mail Code 130-33, Pasadena, CA 91125, USA \\ $^5$ Raymond
and Beverly Sackler School of Physics and Astronomy, Tel Aviv
University, Tel Aviv 69978, Israel \\ $^6$ Universit\"at Z\"urich,
Institut f\"ur Theoretische Physik, Winterthurerstrasse 190, CH-8057
Z\"urich, Switzerland \\ $^7$ Astronomy Centre, Department of Physics
\& Astronomy, University of Sussex, Brighton BN1 9QH, UK\\ $^8$
Department of Astronomy, AlbaNova University Center, Stockholm
University, SE 10691 Stockholm, Sweden \\ $^9$ Department of
Astronomy, University of Texas, Austin, TX 78712-1083, USA ; \\ Texas
Cosmology Center, The University of Texas at Austin, TX 78712, USA}

\begin{document}

\pagerange{\pageref{firstpage}--\pageref{lastpage}} \pubyear{2009}
\maketitle
\label{firstpage}
\begin{abstract}
The 21-cm PDF (i.e., distribution of pixel brightness temperatures) is
expected to be highly non-Gaussian during reionization and to provide
important information on the distribution of density and
ionization. We measure the 21-cm PDF as a function of redshift in a
large simulation of cosmic reionization and propose a simple empirical
fit. Guided by the simulated PDF, we then carry out a maximum
likelihood analysis of the ability of upcoming experiments to measure
the shape of the 21-cm PDF and derive from it the cosmic reionization
history. Under the strongest assumptions, we find that upcoming
experiments can measure the reionization history in the mid to late
stages of reionization to $1-10\%$ accuracy. Under a more flexible
approach that allows for four free parameters at each redshift, a
similar accuracy requires the lower noise levels of second-generation
21-cm experiments.
\end{abstract}


\section{Introduction}\label{intro}

The earliest generations of stars are thought to have transformed the
universe from darkness to light and to have reionized and heated the
intergalactic medium \citep{Review}. Knowing how the reionization
process happened is a primary goal of cosmologists, because this would
tell us when the early stars formed and in what kinds of galaxies. The
clustering of these galaxies is particularly interesting since it is
driven by large-scale density fluctuations in the dark matter
\citep{BLflucts}. While the distribution of neutral hydrogen during
reionization can in principle be measured from maps of 21-cm emission
by neutral hydrogen, upcoming experiments are expected to be able to
detect ionization fluctuations only statistically \citep[for reviews
see, e.g.,][]{fob06,bl07}. Current observational efforts include the
Murchison Widefield Array (MWA, www.haystack.mit.edu/ast/arrays/mwa/),
the Low Frequency Array (www.lofar.org), the Giant Metrewave Radio
Telescope (gmrt.ncra.tifr.res.in), and the Precision Array to Probe
the Epoch of Reionization (astro.berkeley.edu/$\sim$dbacker/eor/).

Studies of statistics of the 21-cm fluctuations have focused on the
two-point correlation function (or power spectrum) of the 21-cm
brightness temperature. This is true both for analytical and numerical
studies and analyses of the expected sensitivity of the new
experiments \citep{miguel,CfA}. The power spectrum is the natural
statistic at very high redshifts, as it contains all the available
statistical information as long as Gaussian primordial density
fluctuations drive the 21-cm fluctuations. More generally, the power
spectrum is also closely related to the directly observed radio
visibilities. Now, during reionization the hydrogen distribution is a
highly non-linear function of the distribution of the underlying
ionizing sources. This follows most simply from the fact that the H~I
fraction is constrained to vary between 0 and 1, and this range is
fully covered in any scenario driven by stars, in which the
intergalactic medium is sharply divided between H~I and H~II regions.
The resulting non-Gaussianity
\citep{Bha} raises the possibility of using complementary statistics to
measuring additional information that is not directly derivable from
the power spectrum \citep[e.g.,][]{fzh04,Ali}.

Numerical simulations have recently begun to reach the large scales
(of order 100 Mpc) needed to capture the evolution of the
intergalactic medium (IGM) during reionization \citep{iliev06,
mellema, zahn, santos}. These simulations account accurately for
gravitational evolution and the radiative transfer of ionizing
photons, but still crudely for gas dynamics and star
formation. Analytically,
\citet{fzh04} used the statistics of a random walk with a linear
barrier to model the H~II bubble size distribution during the
reionization epoch.  Schematic approximations were developed for the
two-point correlation function \citep{fzh04,m05}, but recently
\citet{b07} developed an accurate, self-consistent analytical
expression for the full two-point distribution within the
\citet{fzh04} model, and in particular used it to calculate the 21-cm
correlation function.

Noting the expected non-Gaussianity and the importance of additional
statistics, \citet{fzh04} also calculated the one-point probability
distribution function (PDF) of the 21-cm brightness temperature at a
point. The PDF has begun to be explored in numerical simulations as
well \citep{ciardi,mellema}. Some of the additional information
available in the PDF can be captured by its skewness
\citep{wyithe,Harker}. \citet{DiffPDF} have also considered the 
difference PDF, a two-dimensional function that generalizes both the
one-point PDF and the correlation function and yields additional
information beyond those statistics.

Recently, \citet{Oh} have quantitatively considered the ability of
upcoming experiments to determine the cosmic reionization history from
maximum likelihood fitting of the 21-cm PDF. They specifically used
mixture modeling of the PDF. In this paper we develop a method for
statistical analysis of the PDF that is simpler and more efficient
(allowing, in particular, binning of the PDF). We use our method to
present a quantitative analysis of whether upcoming and future
experiments can measure the detailed shape of the 21-cm PDF and derive
from it the cosmic reionization history. In section~2 we develop our
basic statistical method for fitting the 21-cm PDF, and test it on a
simple toy model for the PDF. We then measure and follow the evolution
of the PDF in a large N-body and radiative transfer simulation of
cosmic reionization; since previous analytical models of the PDF
differ qualitatively from the PDF in the simulation, here we simply
fit the simulated PDF with an empirical, four-parameter model
(section~3). Finally, we present the expected accuracy of
reconstructing the 21-cm PDF and the cosmic reionization history based
on the simulated PDF, either with strict assumptions that lead to one
free parameter at each redshift (section~4), or with a more flexible
approach that allows for four free parameters (section~5). We
summarize our conclusions in section~6.

\section{Basic Method}

In this section we develop our basic statistical method for fitting
the PDF. While the statistical approach is general, for concreteness
we develop it within the context of a simple toy model for the PDF. We
also use this toy, double-Gaussian model in order to get a crude
quantitative intuition on how hard it is to measure the 21-cm PDF. We
note that we follow to some degree \citet{Oh}, who considered such a
double-Gaussian toy model and made a signal-to-noise study of this
model with their analysis method.

\subsection{A Toy Model for the PDF}

\label{s:toy}

It is useful to have a simple PDF example on which to develop and test
our methods. We present here a simplified toy model that captures the
main qualitative features of the PDF as seen in the simulations (and
shown later in the paper) during the central stage of reionization,
when the cosmic ionization fraction $\bar{x}_i \sim 0.3 - 0.6$. The
PDF at this stage has a sharp peak at a differential brightness
temperature (defined as the difference between the actual brightness
temperature and the temperature of the cosmic microwave background at
the same frequency) of $T_b = 0$ mK corresponding to fully ionized
pixels, and another peak at $T_b \sim 20$ mK corresponding to mostly
neutral pixels, with a rapidly declining probability at values above
20 mK, and a smooth probability density in between the peaks that is
lower than the height of either peak. In the observations, this
physical PDF is convolved with a broad Gaussian corresponding to the
thermal noise, resulting in both positive and negative values of
$T_b$. In the limit when we approximate both peaks as delta functions
and neglect the physical PDF at other points, the observed PDF becomes
a sum of two Gaussians with equal standard deviations $\sigma$. While
certainly highly simplified, this model does capture the main question
(relevant especially for low signal-to-noise data, i.e., when $\sigma
\gg 20$ mK) of whether it is at all possible to tell apart the two
peaks and not confuse them with a convolved single peak (i.e., a
single Gaussian).

Thus, we consider two Gaussian distributions with equal standard
deviation $\sigma$ (where $\sigma$ represents the measurement noise
level). In the toy model we use a dimensionless $s$ as the dependent
variable (which represents $T_b$ in the real PDF). The Gaussian
representing the reionized pixels is centered at $s=0$, while the
neutral pixels are represented by a Gaussian centered at $s =
s_G$. The fraction of the total probability contained in the first
Gaussian is $\alpha$. The total distribution is therefore
\beq
p(s) = \alpha G(s, \sigma) + (1-\alpha) G(s-s_G, \sigma)\ ,
\label{eq:toy}
\eeq
where 
\beq G(x, \sigma) \equiv \frac{1}{\sqrt{2\pi} \sigma}
\exp(-x^2/2\sigma^2)\ . \label{eq:G} \eeq
Since in the real case only differences in $T_b$ can be measured, and
not the absolute $T_b$ (which is dominated by foregrounds), in the toy
model we assume that the absolute $s$ cannot be measured. A simple
practical way to do this is to always measure $s$ with respect to its
average value according to the PDF of $s$; we do this separately in
each model and in each simulated data set, and thus only compare the
relative distributions between each model and each data set.

\subsection{Maximum Likelihood and the $C$-Statistic}

In this subsection we develop our basic statistical method for fitting
the PDF, referring to the above toy model as an example for the PDF.
In general, we can create mock data sets by randomly generating $N_p$
values of $s$ from a given $p(s)$ distribution, and we can then try to
estimate the best-fitting parameters with a maximum likelihood
method. For a given mock observed PDF, as given by the $N_p$ generated
values of $s$, we wish to find the best-fitting model PDF $p(s)$ by
maximizing the likelihood $\mathcal{L}$ that the $N_p$ values $s_i$
($i=1$, 2, $\ldots,$ $N_p$) came from $p(s)$. Since the different
values $s_i$ are independent, this probability (apart from fixed
$\Delta s$ factors) is simply
\beq 
\mathcal{L} = \prod_{i=1}^{N_p} p(s_i)\ .
\eeq
Now, it is standard to replace the problem of maximizing the
likelihood $\mathcal{L}$ with a minimization of $-2 \ln \mathcal{L}$,
which in this case is
\beq \label{eq:L1}
-2 \ln \mathcal{L} = -2 \sum_{i=1}^{N_p} \ln p(s_i)\ .
\eeq

In comparing the data to a potential model, we bin the values of $s$
in order to have a manageable number of bins ($N_B=1000$) even when
$N_p$ is very large. This is justified as long as the bin width is
much smaller than any $s$-scale that we hope to resolve in the PDF. We
have explicitly checked that using $N_B=1000$ bins (with the
$C$-statistic, see below) gives the same results as applying
equation~(\ref{eq:L1}) directly, even for the largest values of $N_p$
that we use in this paper. Now, when the expected (according to a
model $p(s)$) number of points $n_{{\rm exp}, j}$ in each bin $j$ is
large (i.e., $n_{{\rm exp}, j} \gg 1$), the actual number $n_j$ has a
standard error of $\sqrt{n_{{\rm exp}, j}}$, and we can find the
best-fitting model by minimizing a standard $\chi^2$ statistic:
\beq \label{eq:chi2}
\chi^2 = \sum_{j=1}^{N_B} \frac{(n_j - n_{{\rm exp}, j})^2} 
{n_{{\rm exp}, j}}\ .\eeq 

However, in modeling the PDF we often wish to include a wide range of
$s$, including some bins where the model probability density $p(s)$ is
very low. When $n_{{\rm exp}, j}$ is small, the $\chi^2$ distribution
with its assumption of a Gaussian distribution for each $n_j$ severely
underestimates the fluctuations in $n_j$ compared to the correct
Poisson distribution. Thus, equation~(\ref{eq:chi2}) can lead to major
errors if $n_{{\rm exp}, j} \ll 1$ in any bin. In this situation, the
correct statistic to use is the $C$-statistic \citep{Cash}, derived
from the Poisson distribution just as the $\chi^2$ statistic is
derived from the Gaussian distribution. The $C$-statistic is defined
as
\beq \label{eq:C}
C = 2 \sum_{j=1}^{N_B} \left( n_{{\rm exp}, j} - n_j \ln n_{{\rm exp},
j}
\right)\ .
\eeq
Note that the Poisson distribution also has a factor of $n_j!$ in the
denominator, which results in an additional $\ln n_j!$ term within the
sum in equation~(\ref{eq:C}), but this term does not depend on the
model parameters (which enter only through $n_{{\rm exp}, j}$) and can
thus be dropped from the minimization.

\subsection{Results for the Toy Model}
\label{s:toy3}

For the toy, double-Gaussian model, the parameters we wish to fit to
mock data sets are $s_G$ and $\alpha$. Note that we assume that
$\sigma$ is known, as we expect that the level of thermal noise per
pixel will be known in the 21-cm experiments, given the known array
properties and the measured foreground level. We perform 1000 Monte
Carlo for each input model, and thus obtain the full distribution of
reconstructed model parameters. In order to develop intuition on how
hard it is to measure the PDF, we define a parameter $\eta$ that
captures a simplistic notion of the total signal-to-noise ratio:
\beq
\eta \equiv \left( \frac{s_G}{\sigma} \right) \sqrt{N_p}\ ,
\label{eq:StN}
\eeq
motivated by $s_G$ as a measure for the signal and $\sigma/\sqrt{N_p}$
as a measure for the effective noise after $N_p$ measurements with
noise $\sigma$ in each. Of course, the ability to detect the two
separate peaks also depends on $\alpha$, in that values close to 0 or
1 make one of the peaks insignificant. For a fixed $\alpha$, though,
we might naively expect that the accuracy of the reconstructed values
of $s_G$ and $\alpha$ would not change with the input value of $s_G$,
as long as we change $N_p$ so as to keep the combination $\eta$ fixed.

\begin{figure}
\includegraphics[width=84mm]{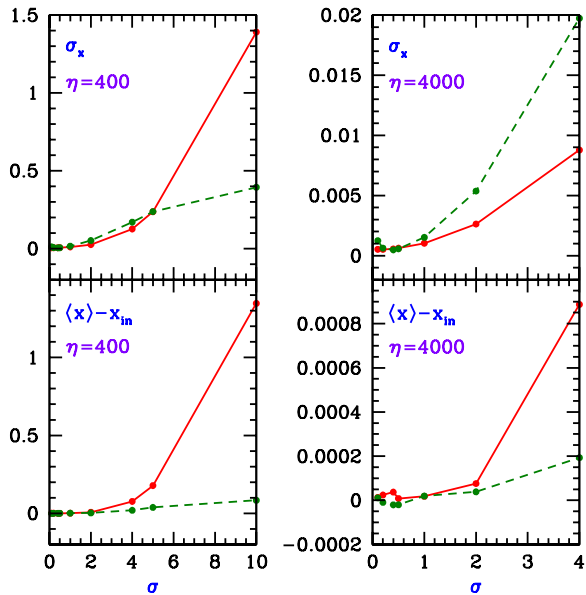}  
\caption{For each model parameter $x$ reconstructed in each Monte 
Carlo trial, we show the bias in the average (i.e., the ensemble average
$\langle x \rangle$ minus the input value $x_{\rm in}$) and the
standard deviation $\sigma_x = \sqrt{\langle x^2 \rangle - \langle x
\rangle^2}$. We consider the model parameters $s_G$ (solid curves, 
input value 1) and $\alpha$ (dashed curves, input value 0.4), as a
function of the noise level (i.e., width of each Gaussian) $\sigma$,
where $\eta$ is held fixed at 400 (left panels) or 4000 (right
panels).}
\label{fig:sigma_ave_sd}
\end{figure}

To test this, we fix the input $\alpha = 0.4$ and $s_G = 1$, and vary
$\sigma$ and $N_p$ together so as to keep $\eta$ fixed. We test
$\eta=400$ and 4000, values comparable to those expected in the real
experiments discussed later in the paper. The Monte Carlo results are
summarized in Figure~\ref{fig:sigma_ave_sd}. The results show that the
parameters can be accurately reconstructed as long as the
signal-to-noise per sample (or per pixel in real data) $s_G/\sigma >
1$. As long as this is the case, the relative error in $s_G$ and
$\alpha$ is no worse than $4\%$ ($\eta=400$) or $0.4\%$ ($\eta=4000$),
and the average reconstructed values are essentially
unbiased. However, once $s_G/\sigma$ drops below unity (i.e., $\sigma
> 1$ in this particular case), the errors increase rapidly with
$\sigma$, so that for $\eta=400$ reconstruction is impossible when
$\sigma=10$ (i.e., both the bias and spread are of order unity) , and
for $\eta=4000$ the errors increase when $\sigma=4$ to a $5\%$
relative spread in $\alpha$.

\begin{figure}
\includegraphics[width=84mm]{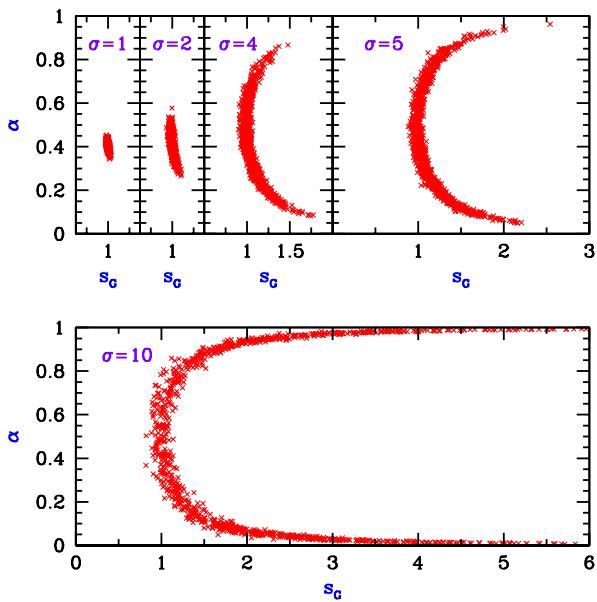}  
\caption{Distribution of reconstructed model parameters $s_G$ and $\alpha$ in 
1000 Monte Carlo simulations. The input parameter values are $s_G=1$
and $\alpha=0.4$. We vary $\sigma$ keeping $\eta = 400$ fixed, so that
the number of samples is $N_p = 160,000\, \sigma^2$. Different panels
cover different $x$ ranges, but all $x$ axes are shown on the same
scale for easy comparison. In the $\sigma \le 2$ panels, small tick
marks are at 0.75 and 1.25.}
\label{fig:dels_chi_MC}
\end{figure}

The reason for these increasing errors is parameter degeneracy, as
illustrated in Figure~\ref{fig:dels_chi_MC} for $\eta=400$. While for
$\sigma=1$ the reconstructed parameter distribution is fairly
symmetrical about the input values of $s_G$ and $\alpha$, resembling a
standard error ellipse, larger $\sigma$ values produce a stretched
error contour that displays a clear (partial) degeneracy between the
parameters $s_G$ and $\alpha$. Intuitively, when $s_G/\sigma \ll 1$
the PDF consists of a narrow input signal (two peaks separated by
$s_G$, in the case of the toy model) convolved with a broad Gaussian
of width $\sigma$. The result is a broad Gaussian of width $\sigma$,
with small bumps (distortions). Apparently these small bumps can be
produced with very different parameter combinations, resulting in a
degeneracy that leads to a large uncertainty when fitting
models. While we have considered here a simple toy model, a similar
degeneracy is encountered with the real 21-cm PDF, as discussed below.

\section{The 21-cm PDF in Simulations}

\subsection{Numerical Simulation}

In this paper we utilize a large-scale N-body and radiative transfer
simulation of cosmic reionization following the methodology first
presented in \citet{iliev06}. The cosmological structure formation and
evolution is followed with a particle-mesh N-body code called PMFAST
\citep{Merz}. These N-body results then provide the
evolving density field of the IGM, as well as the location and mass of
all the halo sources, as input to a separate radiative transfer
simulation of inhomogeneous reionization. The latter is performed with
the ${\rm C}^2-$Ray ({\bf C}onservative, {\bf C}ausal {\bf
Ray}-Tracing) code, a regular-grid, ray-tracing, radiative transfer
and nonequilibrium chemistry code
\citep{methodpaper}. The ionizing radiation is ray-traced from every 
source cell to every grid cell at a given timestep using a method of
short characteristics. ${\rm C}^2-$Ray is designed to be explicitly
photon-conserving in both space and time, which ensures an accurate
tracking of ionization fronts, independently of the spatial and time
resolution. This is true even for grid cells which are very optically
thick to ionizing photons and time steps long compared to the
ionization time of the atoms, which results in high efficiency. The
code has been tested against analytical solutions \citep{methodpaper},
and directly compared with other radiative transfer methods on a
standardized set of benchmark problems
\citep{comparison1,comparison2}.

We simulated the $\Lambda$CDM universe with $1624^3$ dark matter
particles of mass $2.2\times10^7 \,{\rm M}_{\odot}$, in a comoving
simulation volume of $(100\, h^{-1}\,{\rm Mpc})^3$. This allowed us to
resolve (with 100 particles or more per halo) all halos of mass
$2.2\times10^9 \,{\rm M}_{\odot}$ and above. The radiative transfer
grid has $203^3$ cells. The H-ionizing photon luminosities per halo in
our cosmic reionization simulations are assigned as follows. A halo of
mass $M$ is assumed to have converted a mass
$M\cdot(\Omega_{b}/\Omega_{m})\cdot f_{*}$ into stars, where $f_{*}$
is the star formation efficiency. Halo catalogs are discrete in time,
because N-body density fields are stored every $\Delta t\sim20\,{\rm
Myrs}$ and the corresponding halo catalogs are produced at the same
time. If each source forms stars over a period of time $\Delta t$ and
each stellar nucleus\footnote{Note that we defined this number per
atomic nucleus rather than per baryon in stars.} produces $N_{i}$
ionizing photons per stellar lifetime and is used only once per
$\Delta t$, and if a fraction $f_{\rm esc}$ of these photons escape
into the IGM, then the ionizing photon number luminosity of a halo of
mass $M$ is given by
\begin{equation}
Q_{i}=\frac{N_{i}\cdot f_{{\rm esc}}\cdot f_{*}\cdot
M\left(\Omega_{b}/\Omega_{m} \right)}{\Delta t\cdot\mu m_{H}}\ ,
\label{eq:Qi}
\end{equation}
where $m_{H}$ is the mass of a hydrogen atom and $\mu=1.22$ so that
$\mu m_H$ is the mean mass per nucleus. In this model, stars are
produced in a burst, and they keep radiating with a fixed $Q_{i}$ for
$\Delta t\simeq20\,{\rm Myrs}$. We choose here a specific case, first
presented (and labeled f250) in \citet{MN376} and further discussed in
\citet{iliev}. In this scenario, halos are assumed to
host relatively low efficiency emitters, with $f_{\gamma}\equiv
f_{*}f_{{\rm esc}}N_{{\rm i}}=250$ (corresponding, e.g., to Pop II
stars with a Salpeter IMF).

The simulation we use in this work assumes a flat ($\Omega_k=0$)
$\Lambda$CDM cosmology. The simulation is based on the WMAP 3-year
results, which derived the parameters
($\Omega_m,\Omega_\Lambda,\Omega_b,h,\sigma_8,n)=
(0.24,0.76,0.042,0.73,0.74,0.95)$ \citep{ApJS170}.  Here $\Omega_m$,
$\Omega_\Lambda$, and $\Omega_b$ are the total matter, vacuum, and
baryonic densities in units of the critical density, $\sigma_8$ is the
root-mean-square density fluctuation on the scale of $8 h^{-1}{\rm
Mpc}$ linearly extrapolated to the present, and $n$ is the power-law
index of the primordial power spectrum of density fluctuations.

\subsection{The Simulated 21-cm PDF}

During cosmic reionization, we assume that there are sufficient
radiation backgrounds of X-rays and of Ly$\alpha$ photons so that the
cosmic gas has been heated to well above the cosmic microwave
background temperature and the 21-cm level occupations have come into
equilibrium with the gas temperature. In this case, the observed 21-cm
differential brightness temperature (i.e., relative to the cosmic
microwave background) is independent of the spin temperature and, for
our assumed cosmological parameters, is given by
\citep{Madau} \beq T_b = \tilde{T}_b \Psi;\ \ \ \ \tilde{T}_b = 23.7
\left( \frac{\Omega_b h} {0.032} \right) \sqrt{\left( \frac{0.3} 
{\Omega_m} \right) \left( \frac{1+z} {8} \right)}\, {\rm mK}\ ,
\label{eq:Tb}\eeq with $\Psi = x^n [1+\delta]$, where $x^n$
is the neutral hydrogen fraction and $\delta$ is the relative density
fluctuation. Under these conditions, the 21-cm fluctuations are thus
determined by fluctuations in $\Psi$. We denote the PDF by $p(T_b)$,
normalized so that $\int p(T_b) d T_b = 1$.

To calculate the 21-cm PDF, we smooth the 21-cm emission intensity
over our full simulation volume with a cubical top-hat filter
(sometimes referred to as ``boxcar'' averaging) of a pre-determined
size $R_{\rm pix}$. We then assemble the PDF of the resulting values
over a fine grid, much finer than $R_{\rm pix}$. This effectively
smooths out the fluctuations in the PDF and yields a smooth function,
but we note that there is still a real sample (or ``cosmic'') variance
limit on the accuracy of our simulated PDF, resulting from the limited
number of independent volumes of size $R_{\rm pix}$ within our
simulation box. We use $R_{\rm pix} = 5\,h^{-1}$~Mpc,
$10\,h^{-1}$~Mpc, and $20\,h^{-1}$~Mpc, yielding a number of
independent volumes equal to 8000, 1000, and 125, respectively. The
analogous results for the first-year WMAP cosmology were previously
presented, for a few redshifts only, in \citet{mellema} (with a
similarly-defined ionized fraction PDF shown in \citet{iliev06}).

\begin{figure}
\begin{center}
\includegraphics[width=84mm]{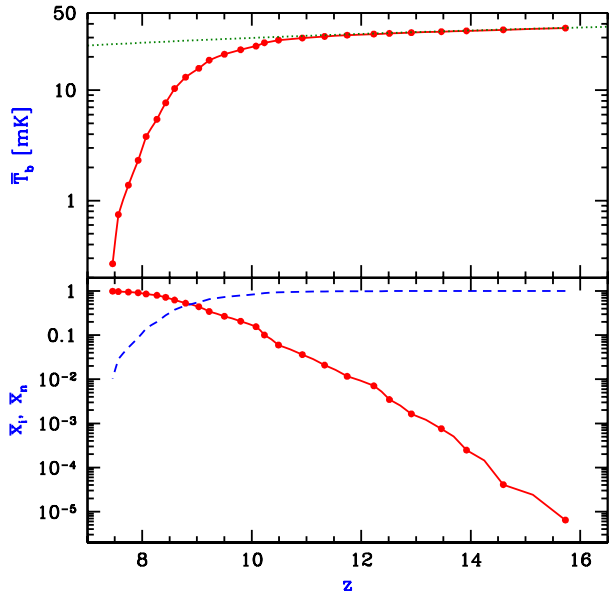} 
\caption{The global progress of cosmic reionization in the simulation,
as a function of the redshift $z$. Bottom panel: we show the
mass-weighted ionized fraction $\bar{x}_i$ (solid curve) and the
corresponding neutral fraction $\bar{x}_n = 1-\bar{x}_i$ (dashed
curve). Top panel: we show the cosmic mean 21-cm differential brightness
temperature $\bar{T}_b$ in the simulation (solid curve), and the mean
$T_b$ expected for a neutral universe of uniform density (dotted
curve). Also indicated in each panel are the 26 output redshifts used
in the analysis below (points).}
\label{fig:z_xi}
\end{center}
\end{figure}

Figure~\ref{fig:z_xi} shows the overall progress of reionization as a
function of redshift in the simulation. We calculate the PDF at 26
redshifts spanning a global mass-weighted ionization fraction
$\bar{x}_i$ from $6 \times 10^{-6}$ to $99.0\%$, with the cosmic mean
21-cm differential brightness temperature $\bar{T}_b$ ranging from
36.5 mK to 0.27 mK. Of course, we assume that $\bar{T}_b$ itself is
not directly observable, due to the bright foregrounds. The main goal
of the PDF analysis is to reconstruct $\bar{x}_i$ vs.\ $z$ using the
$T_b$ fluctuations as captured in the PDF at each redshift.

\begin{figure}
\begin{center}
\includegraphics[width=82mm]{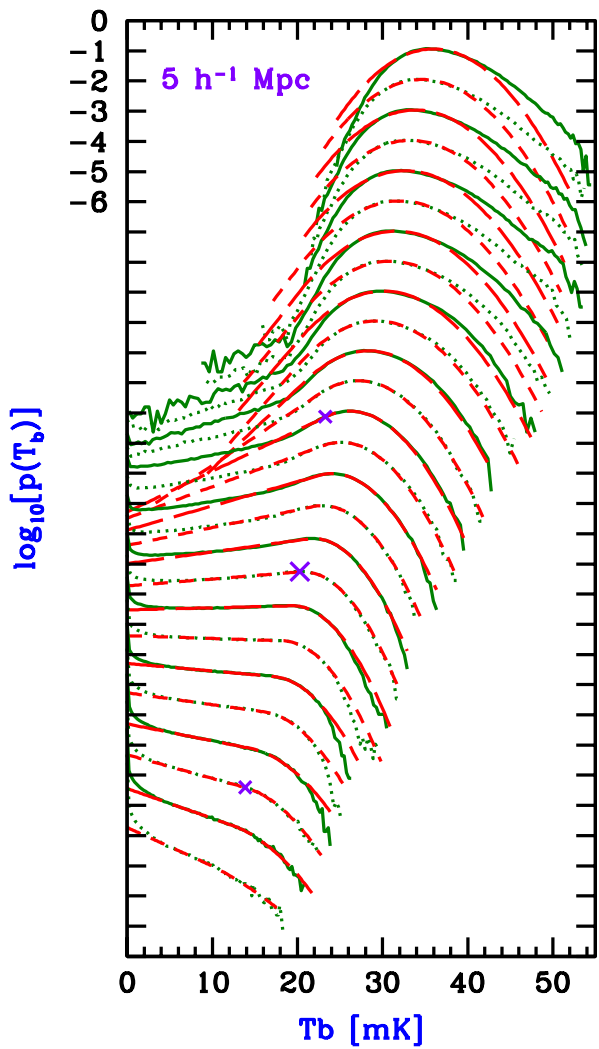} 
\caption{The 21-cm PDF in 5\,$h^{-1}$Mpc cubic pixels, shown versus 
the differential brightness temperature $T_b$. We show $\log_{10}$ of
the PDF, which itself is expressed in units of 1/mK. We show the PDF
obtained from the simulation (alternating solid and dotted curves) and
our best fits to them (alternating long-dashed and short-dashed
curves). The 26 redshifts (see Figure~\ref{fig:z_xi}) range from
$z=15.729$ (top) to 7.460 (bottom). The highest-redshift PDF is shown
at its actual value, corresponding to the labels at the top of the
$y$-axis; each subsequent PDF is shifted vertically down by a factor
of 10 in the PDF. The $\times$ mark points (where $T_b$ equals the
best-fit $T_L$) on three simulated PDFs: early in reionization
($z=10.08$, $\bar{x}_i=0.156$), right after the midpoint ($z=8.79$,
$\bar{x}_i=0.530$), and late in reionization ($z=7.75$,
$\bar{x}_i=0.948$); these points mark the 12-redshift range that is
used in the fitting of mock data in the following sections.}
\label{fig:sim_fit_5}
\end{center}
\end{figure}

We show the measured simulation PDFs for various redshifts and $R_{\rm
pix}=5\,h^{-1}$Mpc in Figure~\ref{fig:sim_fit_5}. The PDF starts out
close to Gaussian at high redshift, when the ionized volume is
negligible and the density fluctuations on the scale $R_{\rm pix}$ are
fairly linear and thus give a Gaussian PDF. There is also a clear
skewness, seen particularly in a high-density tail that drops more
slowly with $T_b$ than the Gaussian fit (more on the fitting function
below); this results from the non-linear growth of density
fluctuations.

As reionization gets under way, the high-density tail drops off and
(coincidentally) approaches the Gaussian shape, as high-density pixels
are more likely to be partially or fully ionized and thus have their
$T_b$ reduced. When $\bar{x}_i$ reaches a fraction of a percent, the
still fairly Gaussian PDF develops a significant low-$T_b$ tail which
is roughly exponential (i.e., linear in the plot of log of the
PDF). This tail corresponds to pixels that are substantially ionized,
i.e., where a large fraction of the pixel volume partially overlaps
one or more ionized bubbles. Soon afterward, a significant peak can be
seen near $T_b=0$ mK, corresponding to fully ionized pixels (i.e.,
pixels in which the hydrogen in the IGM has been fully ionized, but
there may remain a small bit of high-density neutral gas). Near the
mid-point of reionization ($\bar{x}_i = 50\%$), there is still a
half-Gaussian peak (at $T_b \sim 20$ mK), i.e., with a Gaussian
drop-off towards higher $T_b$, now with a nearly flat exponential tail
towards lower $T_b$, and a prominent peak at $T_b=0$ mK. The peak at
zero increasingly dominates towards the end of reionization, as most
pixels become fully ionized, but there remains an exponential tail out
to higher $T_b$, with a cutoff (at $T_b \sim 20$ mK).

\begin{figure}
\begin{center}
\includegraphics[width=84mm]{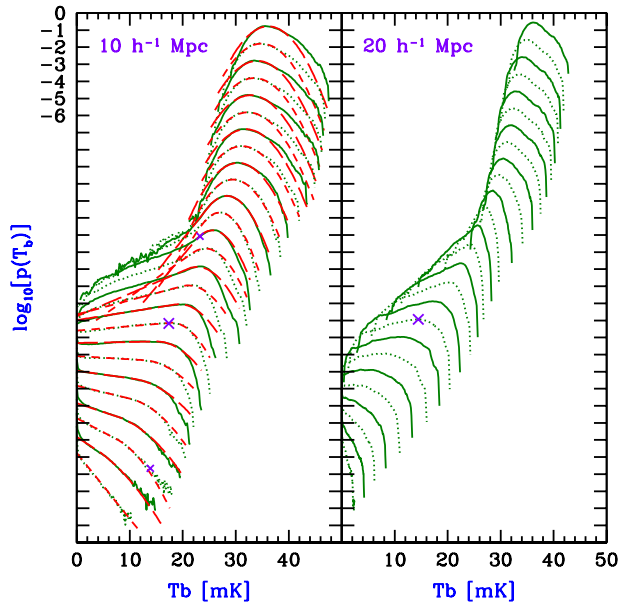} 
\caption{Same as Figure~\ref{fig:sim_fit_5} but for cubic pixels 
of size 10\,$h^{-1}$Mpc (left panel) or 20\,$h^{-1}$Mpc (right
panel). In the right panel we show only the simulated PDFs, and the
$\times$ marks the peak of the PDF right after the midpoint of
reionization ($z=8.79$).}
\label{fig:sim_fit_10}
\end{center}
\end{figure}

The PDFs are shown for $R_{\rm pix}=10\,h^{-1}$Mpc and $20\,h^{-1}$Mpc
in Figure~\ref{fig:sim_fit_10}. The qualitative evolution of the PDF
throughout reionization is similar to the $R_{\rm pix}=5\,h^{-1}$Mpc
case, but the PDF is narrower for larger $R_{\rm pix}$ since the 21-cm
fluctuations are smaller when smoothed on larger scales. Also, for
larger $R_{\rm pix}$ there are fewer pixels in the peak near $T_b=0$
mK since it is more difficult to fully ionize large pixels. The PDFs
for $R_{\rm pix}=20\,h^{-1}$Mpc are not so reliable, as they are
measured based on only 125 independent volumes. Also, their shapes
differ significantly from the PDFs in the smaller pixels, and so they
cannot be successfully fitted with the same model used for the other
PDFs. Thus, in this paper we focus on the two smaller values of
$R_{\rm pix}$ and only present fits to the corresponding
PDFs. Observations of the PDF are most promising during the central
stage of reionization, when the PDF has two significant,
well-separated peaks rather than a single narrow peak (as is the case
either very early or very late in reionization). This two-peak regime
covers $\bar{x}_i \sim 30 - 90\%$ for $R_{\rm pix}=5\,h^{-1}$Mpc, but
only $\bar{x}_i \sim 75 - 95\%$ for $R_{\rm pix}=10\,h^{-1}$Mpc,
because of the rarity of fully ionized pixels in the latter
case. However, even without a strong peak at zero, the extended nearly
flat (exponential) part of the PDF during reionization helps in
measuring the PDF, as we find below.

\subsection{The GED Model Fit to the Simulated PDF}

\label{s:GED}

Previous analytical models of the PDF do not describe our simulated
PDFs well. While the Gaussian at high redshift and the $T_b=0$ mK
Delta function at the end of reionization are obvious, the precise
shape at intermediate redshifts seems to depend on the precise
topology of the ionized bubbles and the geometry of their overlap with
the cubic pixels. Here we take an empirical approach based on our
numerical simulation. Thus we use a Gaussian + Exponential + Delta
function (GED) model for the PDF $p(T_b)$. The Dirac Delta function is
centered at zero, and is connected with an exponential to the
Gaussian. The model depends on four independent parameters: $T_G$
(center of Gaussian), $\sigma_G$ (width of Gaussian), $c_G$ (height of
Gaussian peak) and $T_L$ (transition point between the exponential and
the Gaussian).  Our GED model is thus:
\begin{equation}
p(T_b)=
\cases{
p_1(T_b) = P_D \delta_D(T_b) + a\, \exp(\lambda T_b)\ , & $0 \le T_b
\le T_L$ \cr p_2(T_b) = \displaystyle{ c_G \exp \left[
-\frac{(T_b-T_G)^2}{2\, \sigma_G^2} \right] }\ , & $T_b>T_L$ \cr }
\label{eq:fit_ftn}
\end{equation}
where $\delta_D(x)$ is the Dirac delta function. The quantities $a$
and $\lambda$ can be expressed in terms of the above four parameters
by requiring the exponential and Gaussian functions to connect
smoothly at $T_b=T_L$. The conditions $p_1(T_L) = p_2(T_L)$ and
$p_1^\prime(T_L) = p_2^\prime(T_L)$ lead to
\begin{eqnarray}
\lambda &=& \frac{T_G -T_L}{\sigma_G^2}\ , \\
a &=& c_G \exp \left[ -\frac{(T_L-T_G)^2}{2\, \sigma_G^2} 
-\lambda T_L \right]\ .
\end{eqnarray}
Also, $P_D$ is determined by the requirement of normalization; the
total integrated probability is unity if $P_D = 1 - P_E - P_G$, where
\begin{eqnarray}
P_E &=& \int_{+ \epsilon}^{T_L} p_1(T_b) dT_b = \frac{a}{\lambda}
\left [ \exp(\lambda T_L) - 1 \right]\ , \\
P_G &=& \int_{T_L}^\infty p_2(T_b) dT_b = c_G \sqrt{\frac{\pi}{2}}
\sigma_G\, {\rm erfc}\left( \frac{T_L - T_G}{\sqrt{2}\, \sigma_G}
\right)\ .
\end{eqnarray}
Note that the parameters $P_D$, $P_E$ and $P_G$ represent the relative
contribution to the total probability from the delta function, the
exponential function, and the Gaussian function, respectively.

Using the GED model, we determine the values of $T_G$, $\sigma_G$,
$c_G$ and $T_L$ as functions of redshift by fitting to the simulation
PDFs for pixels of 5\,$h^{-1}$Mpc and 10\,$h^{-1}$Mpc. In approaching
this fitting, we note that we focus on the main features of the PDF,
and not on the fine details. In particular, we do not worry about
features that contain a small fraction of the total probability, or on
the detailed PDF shape on scales finer than several mK. This is
justified since the observations are difficult, and most likely will
not be sensitive to these fine details, at least in the upcoming 21-cm
experiments. In addition, our simulated PDF may not be reliable in its
fine details, since we are using a single, limited simulated volume,
and more generally, numerical simulations of reionization still lack a
detailed demonstration of convergence.

Thus, we do not try to fit the detailed peak shape at $T_b=0$, but
instead represent the total probability of that region with the Delta
function. In practice we only fit to the data beyond the lowest values
of $T_b$, and then set the Delta function contribution $P_D$ to get
the correct overall normalization. Specifically, for each PDF we first
find $T_{b,h}$ which is the highest value of $T_b$ where $p(T_{b,h}) >
10^{-4}$. We then only fit to the data with $T_b > 5$\,mK, if $T_{b,h}
\ge 20$\,mK, or to the data with $T_b > T_{b,h}/4$, if $T_{b,h} <
20$\,mK. At redshifts where the simulation data do not have a Delta
function feature, i.e., there are no pixels near $T_b=0$, we make a
fit constrained by setting $P_D = 1 - P_E - P_G = 0$; this is the case
at the highest redshifts, namely $z \ge 10.924$ for $R_{\rm
pix}=5\,h^{-1}$Mpc and $z \ge 9.034$ for $R_{\rm pix}=10\,h^{-1}$Mpc.

Our GED model fits are shown along with the PDFs in
Figs.~\ref{fig:sim_fit_5} and \ref{fig:sim_fit_10}. The fits are very
good during the central and late stages of reionization, except for
the detailed shape (which we do not try to fit) of the $T_b=0$ peak
which extends out to $T_b \sim 2 - 4$ mK. These are the redshifts that
we focus on in this paper, and which are most promising to observe.
The fits are also quite good at the highest redshifts, where the
simulated PDF is essentially Gaussian except for the skewness. This
skewness, though, affects mainly the tails of the distribution; e.g.,
at the highest redshift ($z=15.729$) for $R_{\rm pix}=5\,h^{-1}$Mpc,
$\sim 60\%$ of the total probability is contained at $T_b$ values
above the peak of the PDF, i.e., the high-density tail adds about
$10\%$ to the $50\%$ of a symmetrical Gaussian. As noted above, this
high-density tail declines with time due to ionization offsetting the
high density of overdense pixels. Thus, the high-$T_b$ tail becomes
well fitted by the Gaussian model once $\bar{x}_i$ rises above a few
percent. At later times the cutoff becomes somewhat steeper than the
Gaussian fit, especially for $R_{\rm pix}=10\,h^{-1}$Mpc, but this
only affects the insignificant tail end of the PDF at the highest
$T_b$. For instance, for $R_{\rm pix}=10\,h^{-1}$Mpc at
$\bar{x}_i=0.530$, the tail beyond $T_b=23$ mK (where the cutoff
starts to differ significantly from the fit) contains only $0.2\%$ of
the total probability. 

Another small mismatch occurs when reionization gets significantly
under way but is still fairly early. The transition region from a
near-Gaussian to a near-exponential shape is not well-fit at these
times by our model, and as a result the fit is significantly below the
low-$T_b$, roughly linear (exponential) tail. This mismatch is
significant in the range of $\bar{x}_i$ from a few percent up to $\sim
30\%$, and at these redshifts this exponential tail typically contains
only a few percent of the total probability (up to $10\%$).

\begin{figure}
\includegraphics[width=84mm]{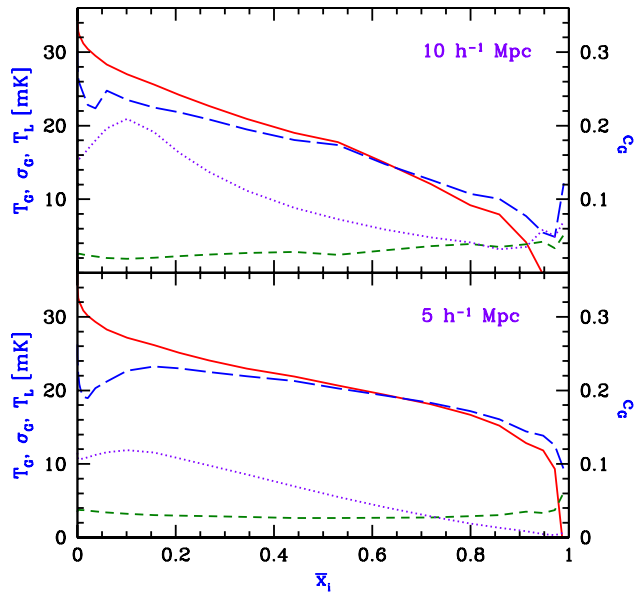}  
\caption{Our best-fitting GED model parameters $T_G$ (solid curve), 
$T_L$ (long-dashed curve), $\sigma_G$ (short-dashed curve), and $c_G$
(dotted curve, different $y$-axis range) as functions of the cosmic
mass-weighted ionization fraction. They are obtained by fitting to the
simulated PDFs for pixels of size 5\,$h^{-1}$Mpc (bottom panel) or
10\,$h^{-1}$Mpc (top panel).}
\label{fig:xi_par}
\end{figure}

Figure~\ref{fig:xi_par} shows how our model parameters vary as cosmic
reionization progresses. The Gaussian peak position $T_G$ and height
$c_G$ both decline with time due to the increasing ionization of even
low-density pixels. At least a half-Gaussian is present until
$\bar{x}_i \sim 60\%$, but after that $T_L > T_G$ and only the
Gaussian cutoff remains. The parameter $\sigma_G$ remains at a value
of a few mK throughout reionization; it gives a measure of density
fluctuations, initially purely and later together with some
correlation with ionization. At the very end of reionization, $T_G
\rightarrow 0$ and then $\sigma_G$ and $c_G$ lose their usual meaning
(e.g., $c_G$ becomes an indirect parameterization of the normalization
of the exponential portion).

\begin{figure}
\includegraphics[width=84mm]{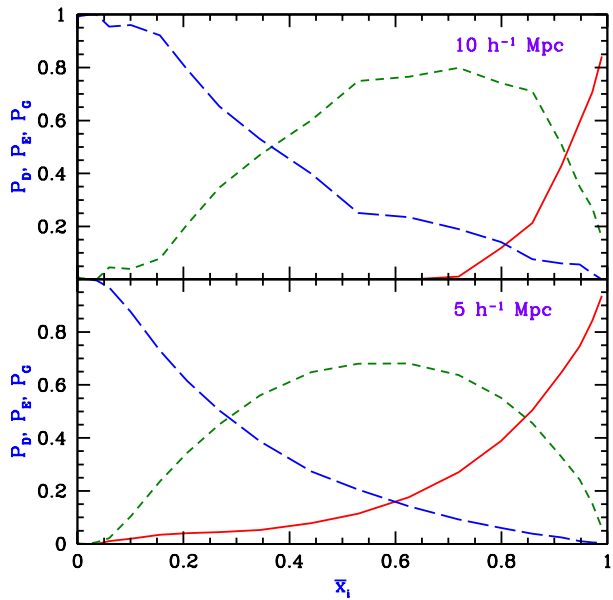} 
\caption{The derived probabilities $P_D$ (solid curve), 
$P_E$ (short-dashed curve) and $P_G$ (long-dashed curve) as functions
of the cosmic mass-weighted ionization fraction. They are obtained by
fitting the GED model to the simulation PDFs for pixels of size
5\,$h^{-1}$Mpc (bottom panel) or 10\,$h^{-1}$Mpc (top panel).}
\label{fig:xi_c}
\end{figure}

Figure~\ref{fig:xi_c} shows the evolution of the probabilities $P_D$
(representing the delta function), $P_E$ (exponential), and $P_G$
(Gaussian), which together add up to unity. The Figure shows how the
21-cm PDF is gradually transformed from a Gaussian to a delta
function, with the exponential dominating at intermediate times (mid
to late reionization). Note that in the limit of infinite resolution,
we would have $P_D=\bar{x}_i$. With a finite resolution, $P_D$ can be
thought of as the cosmic ionized fraction smoothed at the observed
resolution. In practice, converting observed values of $P_D$, $P_E$,
and $P_G$ to the true $\bar{x}_i$ requires some modeling.

\begin{figure}
\includegraphics[width=84mm]{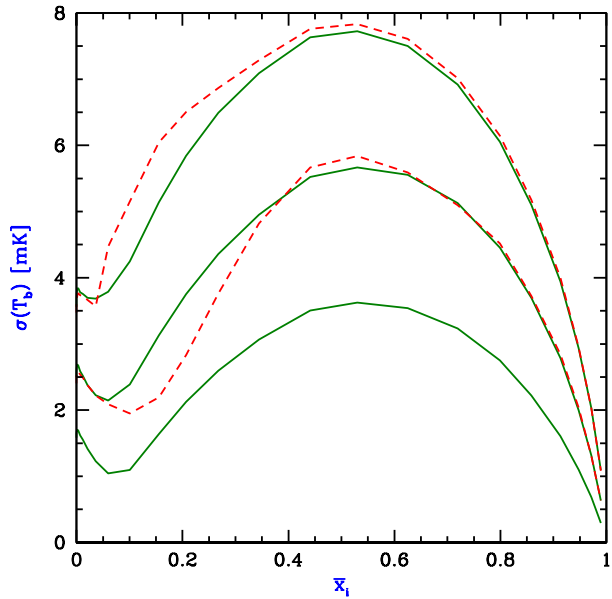} 
\caption{Standard deviation $\sqrt{\langle T_b^2 \rangle - \langle T_b 
\rangle^2}$ as a function of the cosmic mass-weighted ionization fraction. 
We show this quantity for the original simulation data (solid curves)
and from our GED model fits (dashed curves). We consider the PDF in
boxes of size 5\,$h^{-1}$Mpc, 10\,$h^{-1}$Mpc and 20\,$h^{-1}$Mpc (top
to bottom, only simulation data for the 20\,$h^{-1}$Mpc case).}
\label{fig:xim_varTb_all}
\end{figure}

We also calculate the variance $\langle T_b^2 \rangle - \langle T_b
\rangle^2$ from the PDF both directly from the original simulation
data and from our GED model fits. We plot this in
Figure~\ref{fig:xim_varTb_all} for two reasons. First, the plot shows
that the GED model reproduces the variance of the real PDF rather
well, especially where the upcoming measurements are more promising
(i.e., later in reionization). Second, the Figure illustrates a
symmetry in that the variance is maximum near the midpoint of
reionization, and has lower values both before and after the midpoint;
this symmetry helps explain a near-degeneracy that we sometimes find
below, when we consider low signal-to-noise data for which it is
difficult to measure the detailed shape of the PDF, and the variance
is a major part of what can be measured.

\section{Monte-Carlo Results with One Free Parameter}

In the rest of this paper, we present results for the expected
accuracy of reconstructing the 21-cm PDF itself and the cosmic
reionization history from the PDF. To obtain these results, we assume
that our simulation accurately reproduces the real reionization
process in the universe, and furthermore we assume that our GED model
introduced in the previous section can be used as a substitute for the
PDF from the simulation. In the future, more realistic simulations and
more elaborate PDF fits can be used instead, but the general idea will
be the same: as long as the overall signal-to-noise ratio is low, it
is essential to rely on simulations in order to both reconstruct and
interpret the observed PDF.

Of course, even if simulations perfectly predicted the 21-cm PDF for
given inputs, various astrophysical scenarios would give somewhat
different ionizing source and sink properties, and might yield a
variety of possible PDFs. We leave the detailed exploration of this
issue for future work, and here assume that the simulated scenario
matches reality, except that a small number of free parameters are
allowed to vary and must be reconstructed by trying to match the
observed PDF. In this section, we reconstruct reionization from the
PDF under the most optimistic assumption, where we assume that the
real PDF matches the simulated one as a function of just a single
parameter, the ionization fraction $\bar{x}_i$. Thus, at each
redshift, we find the value of $\bar{x}_i$ that best matches the
observed PDF, assuming that the PDF varies with $\bar{x}_i$ as in the
simulation. In practice we expect that $\bar{x}_i$ is indeed the main
parameter that determines the PDF, but there should be some small
additional dependence on redshift and astrophysical inputs. In the
next section we explore a more flexible approach which makes much
weaker assumptions.

Thus, here we wish to know how well a certain experiment can determine
$\bar{x}_i$ assuming this one-parameter model. An experiment is
specified by a total number of pixels $N_p$ and a noise per pixel
$\sigma_N$. We can simulate an observed PDF from such an experiment at
a given input $\bar{x}_i$ by generating $N_p$ data points from the PDF
of equation~(\ref{eq:fit_ftn}) and adding to each noise generated from
a Gaussian distribution with standard deviation $\sigma_N$. The
resulting Monte-Carlo-generated ``observed'' PDF is then compared, via
the $C$-statistic of equation~(\ref{eq:C}), to the model, which is
equation~(\ref{eq:fit_ftn}) convolved with the Gaussian noise. This
convolved function $q(T_b)$ equals:
\begin{eqnarray}
q(T_b) = P_D\, G(T_b,\sigma_N) + q_1(T_b) + q_2(T_b)\ ,
\end{eqnarray}
where $G$ is a Gaussian (eq.~\ref{eq:G}), and 
\begin{eqnarray}
q_1(T_b) &=& \frac{1}{2} a \exp \left( \frac{\lambda^2 \sigma_N^2}{2} +
\lambda T_b \right) \times \\
& & \left\{ {\rm erf}\left( \frac{\lambda \sigma_N^2 +T_b} {\sqrt{2}
\sigma_N} \right) - {\rm erf}\left( \frac{\lambda \sigma_N^2 - T_L
+T_b}{\sqrt{2} \sigma_N} \right) \right\}\ , \nonumber \\ q_2(T_b) &=&
\frac{1}{2} c_G \frac{\sigma_G}{\sigma_c} \exp \left\{
-\frac{(T_b-T_G)^2}{2\sigma_c^2} \right\} \times \\ & & {\rm
erfc}\left\{ \frac{\sigma_N^2(T_L-T_G) + \sigma_G^2 (T_L-T_b)}{\sqrt{2}
\sigma_c \sigma_G \sigma_N} \right\}\ , \nonumber 
\end{eqnarray}
where $\sigma_c^2 = \sigma_G^2 + \sigma_N^2$. As noted above, in this
section we regard $q(T_b)$ as a one-parameter function of $\bar{x}_i$,
taking $T_G$, $\sigma_G$, $c_G$ and $T_L$ to be functions of
$\bar{x}_i$ as shown in Figure~\ref{fig:xi_par}. For clarity we denote
the input, real cosmic ionized fraction simply $\bar{x}_i$, while the
free parameter which is the output of the fitting we denote
$\bar{x}_i^{\rm out}$. Note that we assume that the experimental setup
is sufficiently well characterized that $\sigma_N$ is known and need
not be varied in the fitting. Also note that while the various
temperatures we have defined ($T_b$, $T_L$, and $T_G$) refer to the
differential brightness temperature (i.e., 0~mK refers to the absence
of a cosmological signal), in practice, when the foregrounds as well
as the cosmic microwave background are subtracted, the mean
cosmological signal on the sky will be removed as well, since there is
no easy way to separate out different contributions except through
their fluctuations. Thus, as in section~\ref{s:toy}, we assume that
the absolute $T_b$ cannot be measured, and in our fitting always
measure $T_b$ with respect to its average value according to the PDF,
both in each model and in each simulated data set.

For the experimental specification, we adopt the (rough) expected
parameters for one-year observations of a single field of view with
the MWA. We use the relations for 21-cm arrays from the review by
\citet{fob06}, adopting a net integration time $t_{\rm int} = 1000$
hours, a collecting area $A_{\rm tot}=7 \times 10^3\ {\rm m}^2$, a
field of view of $\pi 16^2\ {\rm deg}^2$, and a total bandwidth
$\Delta \nu_{\rm tot} = 6$~MHz. Then assuming cubic pixels of comoving
size $r_{\rm com}$, we find
\begin{eqnarray}
N_p &=& 6.0 \times 10^6\, \left(\frac{r_{\rm com}}{5 h^{-1} {\rm
Mpc}}\right)^{-3}\,\left(\frac{1+z}{10}\right)^{0.9}\ , \\
\sigma_N &=& 200 \left(\frac{r_{\rm com}} 
{5 h^{-1} {\rm Mpc}}\right)^{-2.5}\,
\left(\frac{1+z}{10}\right)^{5.25}\,{\rm mK}\ .
\end{eqnarray}
In order to explore the dependence on the noise level, we also
consider various specifications with lower noise in the same field of
view, e.g., 1/2 the noise we denote as MWA/2 (which corresponds, e.g.,
to four-year data with the MWA), while 1/10 the noise we denote as
MWA/10 (which corresponds to the regime of larger, second-generation
21-cm arrays). Note that we include only Gaussian thermal noise, whose
magnitude is determined by the receiver's system temperature, which in
turn is set by the sky's brightness temperature which is dominated by
Galactic synchrotron emission \citep{fob06}. In particular this
assumes perfect foreground removal from the 21-cm maps; we leave an
analysis of the effect of foreground residuals for future work.

We note the following conversions between comoving distance and
observational units of angular and frequency resolution:
\beq
5 h^{-1} {\rm Mpc} \approx 2.6^\prime\,\left(\frac{1+z} {10}
\right)^{-0.2} \approx 0.37\, {\rm MHz} \,\left(\frac{1+z} {10}
\right)^{-0.5} \ .
\eeq
The diffraction limit of the MWA is several arcminutes, but its
frequency resolution will be around 10~kHz. In principle, this allows
a measurement of the PDF in skinny boxes (thinner in the redshift
direction) rather than cubes. This would give us more points but with
less signal in each, keeping the overall signal-to-noise ratio about
the same. By accessing fluctuations on smaller scales, this skinny-box
PDF would be somewhat broader than the cubic one but on the other
hand, our quantitative results for the toy model above suggest that
decreasing the signal-to-noise ratio per pixel in this way would have
a strong tendency to introduce partial degeneracies. Thus, we do not
expect this option to be productive (except in the cases when
the errors in the cubic PDF are very small), and focus here on
the simplest case of the 21-cm PDF measured in cubes.

\begin{figure}
\includegraphics[width=84mm]{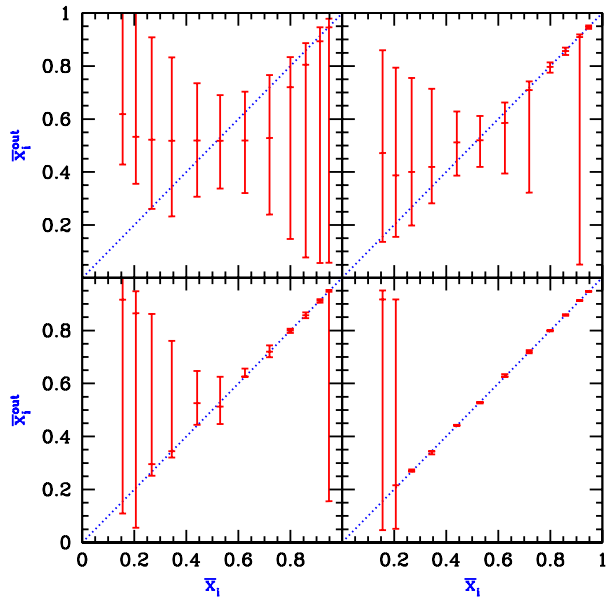} 
\caption{Expected 1$\sigma$ errors on reconstructing the cosmic
mean ionized fraction from the PDF, assuming just one free parameter.
Specifically, for each input value of $\bar{x}_i$ we show the output
median (i.e., 50 percentile) $\bar{x}_i^{\rm out}$ as well as the 16
to 84 percentile range. We consider MWA 1-yr errors (left panels) or
MWA/2 (right panels), for the PDF in 5\,$h^{-1}$Mpc boxes (top panels)
or 10\,$h^{-1}$Mpc boxes (bottom panels).}
\label{fig:xi_MC_MWA}
\end{figure}

At each redshift, we generate 1000 Monte Carlo instances of observed
PDFs and minimize the $C$-statistic to find the best-fitting model in
each case. Results for MWA and MWA/2 errors are plotted in
Figure~\ref{fig:xi_MC_MWA}, which shows that for first-generation
experiments the larger (10\,$h^{-1}$Mpc) boxes are much more
promising, since the lower noise $\sigma_N$ (by a factor of $\sim6$)
dominates despite the narrower PDF (compare Figs.~\ref{fig:sim_fit_5}
and \ref{fig:sim_fit_10}) and smaller number of pixels $N_p$ (by a
factor of 8). We note that lower noise is particularly important in
view of the partial degeneracy (demonstrated in section~\ref{s:toy3}
for the toy model) that arises when $\sigma_N$ is greater than the
characteristic width of the intrinsic PDF. The partial degeneracy is
also apparent in comparing the MWA and MWA/2 cases, where at some
$\bar{x}_i$ values, halving the errors crosses a degeneracy threshold
and cuts the output uncertainty in a non-linear fashion. We caution
that cases that are very near such a threshold may be susceptible to
additional numerical errors.

\begin{figure}
\includegraphics[width=84mm]{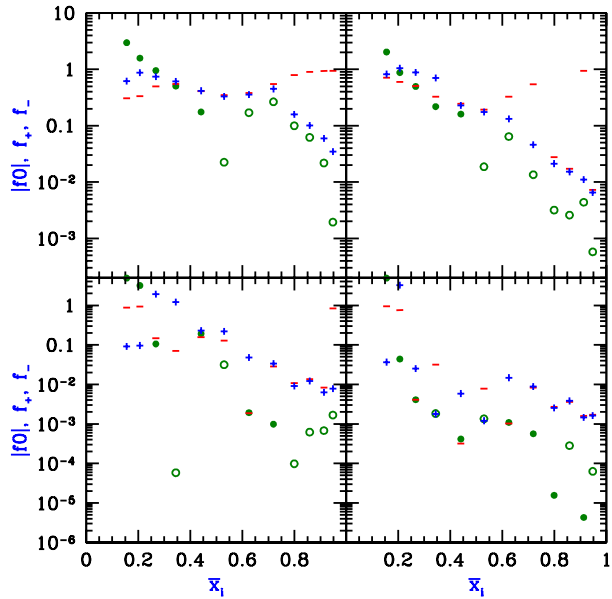} 
\caption{Same as Figure~\ref{fig:xi_MC_MWA} but showing relative errors 
(see text), for better visibility of cases with small errors. We show
$f_0$ (absolute value shown, where negative values are open circles
and positive values are solid circles) , $f_+$ ($+$ symbols), and
$f_-$ ($-$ symbols). We consider MWA 1-yr errors (left panels) or
MWA/2 (right panels), for the PDF in 5\,$h^{-1}$Mpc boxes (top panels)
or 10\,$h^{-1}$Mpc boxes (bottom panels). }
\label{fig:xi_MC_MWAb}
\end{figure}

The same results are shown in Figure~\ref{fig:xi_MC_MWAb} in terms of
relative errors, making it easier to see and compare both small and
large errors. Specifically, in terms of the various percentile output
ionization fractions (e.g., we denote the median by $\bar{x}_i^{\rm
out,50}$), we show $f_0 = (\bar{x}_i^{\rm out,50}/\bar{x}_i) - 1$ (the
relative difference between the median output value and the true input
value, representing the fractional bias of the reconstruction), $f_+ =
(\bar{x}_i^{\rm out,84}/\bar{x}_i^{\rm out,50}) - 1$ (the relative
difference between the $84\%$ and median values, representive the
fractional $+1\sigma$ spread), and $f_- = 1 - (\bar{x}_i^{\rm
out,16}/\bar{x}_i^{\rm out,50})$ (the relative difference between the
$16\%$ and median values, representive the fractional $-1\sigma$
spread). The Figure shows that the reconstruction is typically
unbiased within the errors (i.e., the $1\sigma$ range is significantly
larger than the bias in the median), except for some points in the
early stages of reionization. Only a little information is available
with the PDF in the smaller boxes (except for a few redshifts with
MWA/2 errors); typically the error ranges are smaller near the
mid-point of reionization, partly due to the fact (see
Figure~\ref{fig:xim_varTb_all}) that the variance of the PDF suffices
to distinguish the mid-point of reionization from its two ends, but
the early and late stages are degenerate with each other in terms of
the variance. A rather good measurement of the reionization history is
expected with 10\,$h^{-1}$Mpc boxes, in the mid to late stages of
reionization, down to $1\%$ errors in measuring the cosmic mean
ionized fraction (or even better with MWA/2 errors). When the errors
are small, the measurement is unbiased and has symmetric error bars.

\begin{figure}
\includegraphics[width=84mm]{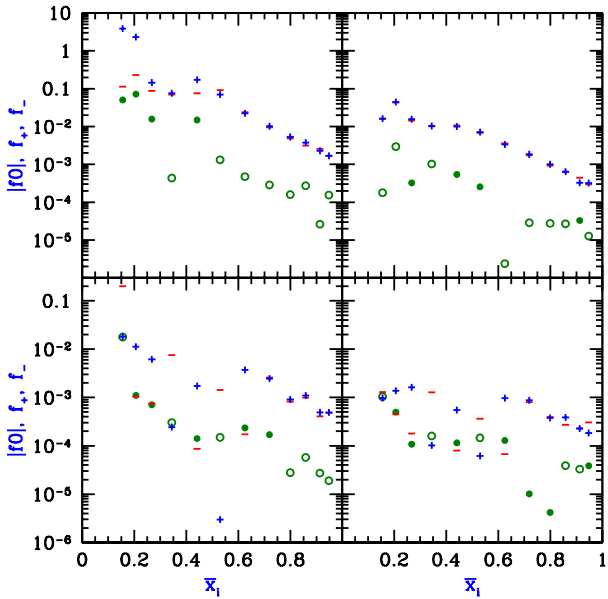} 
\caption{Same as Figure~\ref{fig:xi_MC_MWAb}, but we consider MWA/4
errors (left panels) or MWA/10 (right panels), for the PDF in
5\,$h^{-1}$Mpc boxes (top panels) or 10\,$h^{-1}$Mpc boxes (bottom
panels). }
\label{fig:xi_f4_MWAb}
\end{figure}

As shown in Figure~\ref{fig:xi_f4_MWAb}, lower errors (approaching
second-generation experiments) would avoid the degeneracy and allow a
meaningful measurement of the cosmic reionization history even with
the PDF in the smaller boxes, but 10\,$h^{-1}$Mpc boxes always give a
more precisely measured output value by about an order of magnitude.
The expected success in reconstructing the reionization history under
the strict assumption of a single free parameter motivates us to
consider in the following section a more flexible reconstruction
method.

\section{Monte-Carlo Results with a Flexible Four-Parameter Model}

In the previous section we showed the expected accuracy of
reconstructing the cosmic reionization history from the 21-cm PDF,
assuming the PDF shape is known as a function of the cosmic mean
ionized fraction. In this section we drop the latter assumption and
present results for the expected accuracy of reconstructing the
detailed shape of the 21-cm PDF directly from the data. We focus on
the regime of second-generation experiments, since the expected MWA
errors do not allow such a reconstruction. Even with the lower errors,
the PDF cannot be reconstructed parameter free, so we assume that our
four-parameter GED model from section~\ref{s:GED} correctly describes
the real intrinsic PDF (an assumption which is explicitly true in our
Monte-Carlo setup). Otherwise we do not assume any restrictions, and
allow the four parameters of the model to vary freely when fitting
(again by minimizing the $C$-statistic) to the noisy mock PDF data.

\begin{figure}
\begin{center}
\includegraphics[width=84mm]{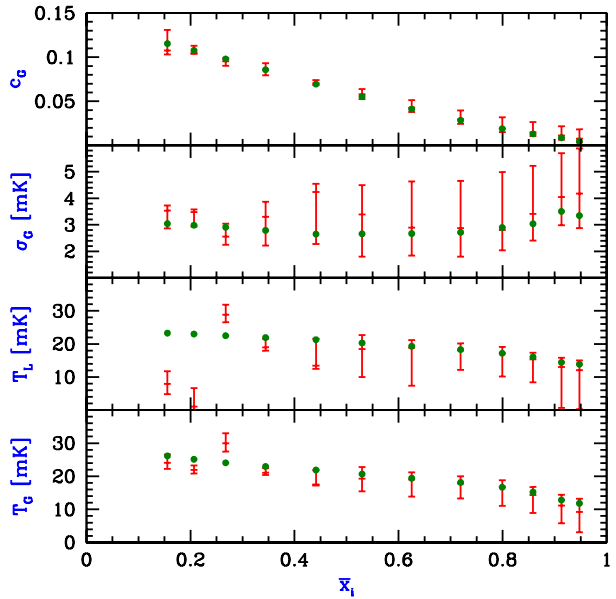}   
\caption{Expected 1$\sigma$ errors on reconstructing the
PDF parameters assuming the four-parameter GED model, assuming MWA/10
errors on the PDF in 5\,$h^{-1}$Mpc boxes. We show the 16, 50, and 84
percentiles, as before, and also the assumed input values (circles).}
\label{fig:4par_f10}
\end{center}
\end{figure}

\begin{figure}
\begin{center}
\includegraphics[width=84mm]{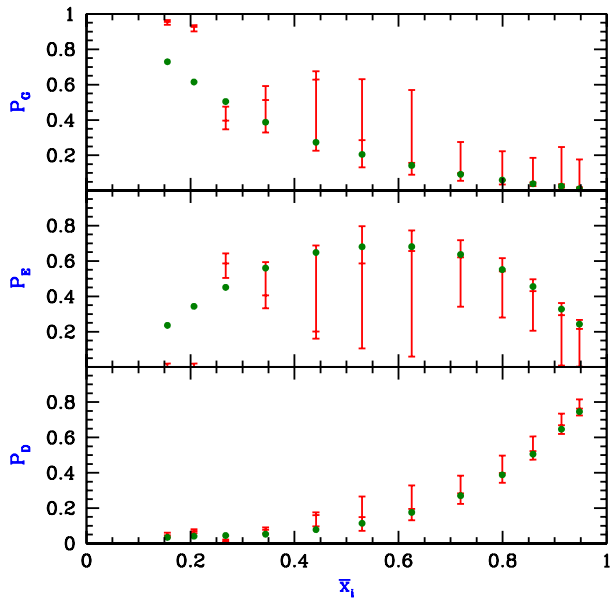}   
\caption{Expected 1$\sigma$ errors on reconstructing the
derived probabilities of the GED model, from a four-parameter fit to
the PDF, assuming MWA/10 errors and 5\,$h^{-1}$Mpc boxes. We show the
16, 50, and 84 percentiles, as before, and also the assumed input
values (circles).}
\label{fig:4par_f10b}
\end{center}
\end{figure}

Specifically, we fit the four parameters $T_G$, $T_L$, $\sigma_G$, and
$c_G$. We consider fitting the PDF in 5\,$h^{-1}$Mpc boxes with MWA/10
or MWA/20 errors. Figure~\ref{fig:4par_f10} shows that significant
information can be reconstructed with MWA/10 errors, although the
errors in the reconstructed parameters are usually fairly large (with
particular failures at the early stage of reionization). The derived
total probabilities of the GED model components are shown in
Figure~\ref{fig:4par_f10b}; in particular, the statistically
significant measurement of the evolution of $P_D$ (which is the cosmic
ionized fraction smoothed over the 5\,$h^{-1}$Mpc resolution) shows
that significant information can be extracted about the cosmic
reionization history, even in this more flexible fitting approach.

\begin{figure}
\begin{center}
\includegraphics[width=84mm]{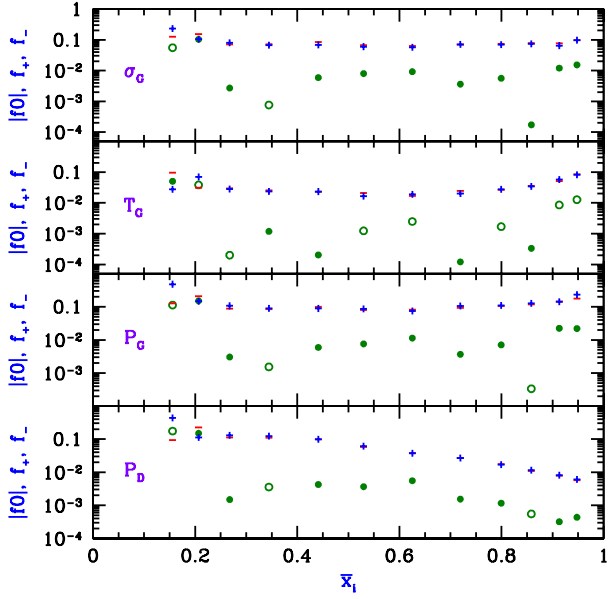}   
\caption{Expected 1$\sigma$ errors on reconstructing various quantities
of the GED model, from a four-parameter fit to the PDF, assuming
MWA/20 errors and 5\,$h^{-1}$Mpc boxes. As in the previous section, we
show the relative errors $f_0$ (absolute value shown, where negative
values are open circles and positive values are solid circles) , $f_+$
($+$ symbols), and $f_-$ ($-$ symbols). }
\label{fig:4par_f20}
\end{center}
\end{figure}

Since the errors on the reconstructed parameters with MWA/10 noise are
still mostly of order unity, we explored further and found that MWA/20
is necessary to break most of the degeneracies.
Figure~\ref{fig:4par_f20} shows that in this case the parameters can
usually be reconstructed to $1-10\%$ accuracy (with symmetric error
bars and insignificant bias). Specifically we show the four quantities
$P_D$, $P_G$, $T_G$ and $\sigma_G$, which together comprise a complete
set that specifies the GED model. Note that the measurement of $P_D$
is particularly precise, during the latter stages of reionization.

\begin{figure}
\begin{center}
\includegraphics[width=84mm]{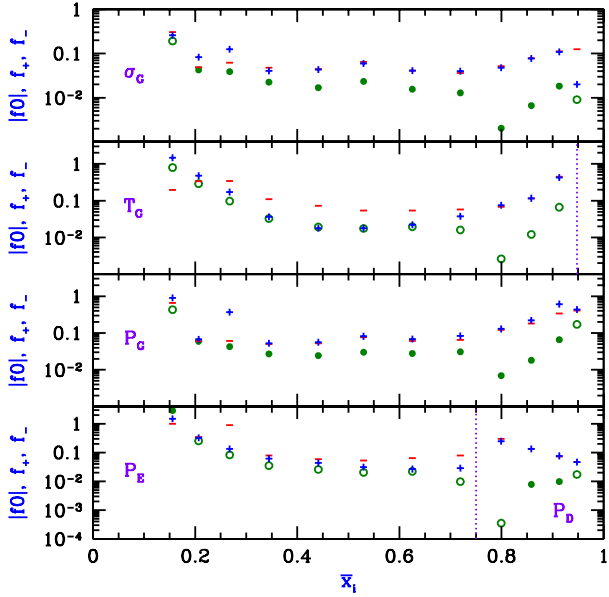}   
\caption{Expected 1$\sigma$ errors on reconstructing various quantities
of the GED model, from a four-parameter fit to the PDF, assuming MWA/5
errors and 10\,$h^{-1}$Mpc boxes. We show the relative errors $f_0$
(absolute value shown, where negative values are open circles and
positive values are solid circles) , $f_+$ ($+$ symbols), and $f_-$
($-$ symbols). }
\label{fig:4par_f5_10}
\end{center}
\end{figure}

\begin{figure}
\begin{center}
\includegraphics[width=84mm]{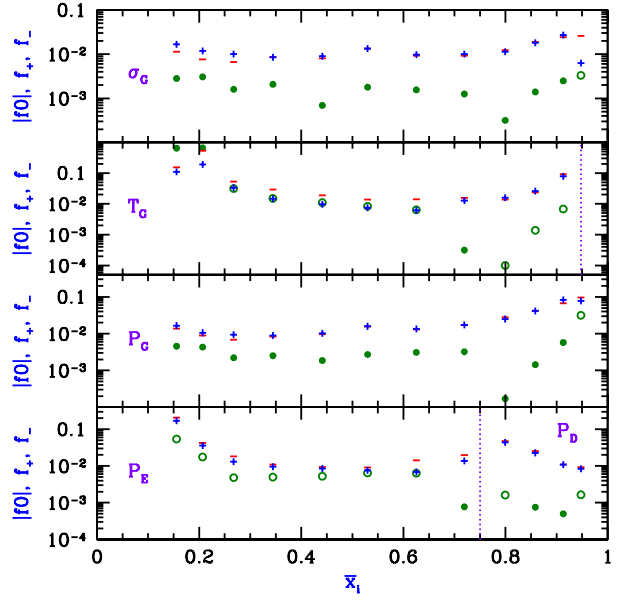}   
\caption{Expected 1$\sigma$ errors on reconstructing various quantities
of the GED model, from a four-parameter fit to the PDF, assuming
MWA/10 errors and 10\,$h^{-1}$Mpc boxes. We show the relative errors
$f_0$ (absolute value shown, where negative values are open circles
and positive values are solid circles) , $f_+$ ($+$ symbols), and
$f_-$ ($-$ symbols). }
\label{fig:4par_f10_10}
\end{center}
\end{figure}

As in the previous section, the PDF in larger, 10\,$h^{-1}$Mpc boxes,
is easier to measure, due to the lower noise per pixel. Thus, here we
consider somewhat larger noise levels, MWA/5 and MWA/10, with results
shown in Figs.~\ref{fig:4par_f5_10} and \ref{fig:4par_f10_10}. Note
that the last (highest $\bar{x}_i$) point in $T_G$ is not shown, since
the input $T_G$ there is zero (see Figure~\ref{fig:xi_par}), and also,
we show $P_D$ only during late reionization, where it is non-zero (see
Figure~\ref{fig:xi_c}), and $P_E$ at earlier times. While the errors
are fairly large with MWA/5 errors, they reach the $1-10\%$ level with
MWA/10, corresponding to a second-generation 21-cm experiment.

\section{Conclusions}

We have carried out a detailed quantitative analysis of whether
upcoming and future experiments can measure the shape of the 21-cm PDF
and derive from it the cosmic reionization history. This is an
important question since the PDF during reionization is highly
non-Gaussian, it directly provides important information such as the
cosmic ionization fraction at each redshift (though smoothed on the
experimental resolution scale), and is potentially a way to derive the
cosmic reionization history independently of the standard power
spectrum analysis.

We developed a maximum-likelihood approach that achieves maximum
efficiency by minimizing the $C$-statistic (eq.~\ref{eq:C}) applied to
binned PDF data. We used a toy PDF model of two Gaussians
(eq.~\ref{eq:toy}) to show that the simplistic notion of
signal-to-noise ratio (eq.~\ref{eq:StN}) does not fully describe the
ability to extract the PDF out of noisy data. Instead, once the noise
per pixel rises above a few times the signal (i.e., the width of the
intrinsic PDF), the errors blow up due to a strong degeneracy, even if
the total signal-to-noise ratio is kept fixed by increasing the number
of pixels (Figs.~\ref{fig:sigma_ave_sd} and \ref{fig:dels_chi_MC}).

We measured the 21-cm PDF as a function of redshift in a large-scale
N-body and radiative transfer simulation of cosmic reionization
(Figs.~\ref{fig:sim_fit_5} and \ref{fig:sim_fit_10}). The PDF starts
out close to Gaussian at high redshift, due to still-linear density
fluctuations, later develops an exponential tail at low $T_b$, and
finally becomes strongly peaked at zero towards the end of
reionization. We empirically fit the PDF from the simulation with a
four-parameter Gaussian + Exponential + Delta function (GED) model
(eq.~\ref{eq:fit_ftn}, Figs.~\ref{fig:xi_par} and \ref{fig:xi_c}).

Assuming the simulations as a reliable guide for the evolution of the
PDF, we quantitatively explored how well parameters can be measured
with two different approaches. In the most optimistic approach, we
assumed that the real PDF matches the simulated one as a function of
just a single free parameter, the ionization fraction $\bar{x}_i$, and
tried to reconstruct this parameter from noisy mock data. We found
that first-generation experiments (such as the MWA) are promising, at
least if relatively large (10\,$h^{-1}$Mpc) pixels are used along with
their relatively low noise level per pixel. Specifically, a rather
good measurement of the reionization history is expected in the mid
to late stages of reionization, down to $1\%$ errors in measuring the
cosmic mean ionized fraction.

We also considered reconstructing the cosmic reionization history
together with the PDF shape, all while assuming that the
four-parameter GED model correctly describes the real intrinsic PDF,
but allowing the four parameters to vary freely when fitting mock data
at each redshift. We found that this flexible approach requires much
lower noise levels, characteristic of second-generation 21-cm
experiments, to reach the level of $1-10\%$ accuracy in measuring the
parameters of the 21-cm PDF.

We note that cosmic reionization ends in our simulation at redshift
7.5 (Fig.~\ref{fig:z_xi}). If reionization in the real universe ends
later (e.g., closer to $z=6.5$), then observations will be somewhat
easier than we have assumed, due to the reduced foregrounds at lower
redshift. On the simulation side, further work is necessary to
establish the numerical convergence of the simulated 21-cm PDF during
reionization, and to explore the dependence of the PDF on various
astrophysical scenarios for the ionizing sources and sinks during
reionization. This further effort is warranted since we have shown
that the 21-cm PDF is a promising alternative to the power spectrum
which can independently probe the cosmic reionization history.

\section*{Acknowledgments}
We thank Oh, Hansen, Furlanetto, \& Mesinger for provided us with a
draft of their paper far in advance of publication. RB is grateful for
support from the ICRR in Tokyo, Japan, the Moore Distinguished Scholar
program at Caltech, the John Simon Guggenheim Memorial Foundation, and
Israel Science Foundation grant 629/05. This study was supported in
part by Swiss National Science Foundation grant 200021-116696/1, NSF
grant AST 0708176, NASA grants NNX07AH09G and NNG04G177G, Chandra
grant SAO TM8-9009X and Swedish Research Council grant 60336701.


\label{lastpage}

\end{document}